\documentclass[article]{siamart171218}

\usepackage[utf8]{inputenc}
\usepackage[T1]{fontenc}
\usepackage{subfigure}
\usepackage{lineno}
\usepackage{amsopn}
\usepackage{lipsum}
\usepackage{amsfonts}
\usepackage{graphicx}
\usepackage{epstopdf}
\usepackage{algorithmic}
\ifpdf
  \DeclareGraphicsExtensions{.eps,.pdf,.png,.jpg}
\else
  \DeclareGraphicsExtensions{.eps}
\fi

\title{Compressive Sampling for Array Cameras}

\author{Xuefei Yan\thanks{Camputer Laboratory, Kunshan China
  (\email{yanxf@ksitri.com}).}
\and David J. Brady\thanks{Department of Electrical and Computer Engineering, Duke University, Durham, NC
  (\email{dbrady@duke.edu})}
\and Jianqiang Wang\thanks{Nanjing University, School of Electronic Science and Engineering, Nanjing, China
  (\email{1139970854@qq.com}).}
\and Chao Huang\thanks{Nanjing University, School of Electronic Science and Engineering, Nanjing, China
  (\email{283753123@qq.com}).}
    \and Zian Li\thanks{Camputer Laboratory, Kunshan China
  (\email{zian\_li@126.com}).}
\and Songsong Yan\thanks{Camputer Laboratory, Kunshan China
  (\email{yanss@ksitri.com}).}
\and Di Liu\thanks{Camputer Laboratory, Kunshan China
  (\email{liudi@ksitri.com}).}
\and Zhan Ma\thanks{Nanjing University, School of Electronic Science and Engineering, Nanjing, China
  (\email{mazhan@nju.edu.cn}).}
  }

\begin{document}

\maketitle

\begin{abstract}
While design of high performance lenses and image sensors has long been the focus of camera development, the size, weight and power of image data processing components is  currently the primary barrier to radical improvements in camera resolution. Here we show that Deep-Learning-Aided Compressive Sampling (DLACS) can reduce operating power on camera-head electronics by 20x. Traditional compressive sampling has to date been primarily applied in the physical sensor layer, we show here that with aid from deep learning algorithms, compressive sampling offers unique power management advantages in digital layer compression.
\end{abstract}



\section{Introduction}

In the 75 year transition from standard definition to 8K, the pixel capacity of video has increased by a factor of ~100. Over the same time period, the computer was invented and the processing, communications and storage capacities of digital systems improved by 6-8 orders of magnitude. The failure of video resolution to develop a rate comparable to other information technologies may be attributed to the physical challenge of creating lenses and sensors capable of capturing more than 10 megapixels. Recently, however, parallel \cite{parallel} and multiscale \cite{nature, multiscale} optical and electronic designs have enabled video capture with resolution in the range of 0.1-10 gigapixels per frame. 
At 10 to 100 gigapixels, video capacity will have increased by a factor comparable to improvements in other information technologies. 

 Since multiscale and array optics have largely resolved lens design challenges associated with gigapixel-scale cameras, the size, weight, power and cost (SWaP) of electronic components capable of processing and storing such video is the primary remaining barrier to gigapixel-scale cameras. Through the AWARE~\cite{Nichols:16} and Mantis~\cite{parallel} programs we know that the size, weight and power of electrical components is currently the primary barrier to compact gigapixel scale cameras. In these systems, image sensors operating at 30 fps draw ~100 milliwatts/megapixel, whereas the image processing pipeline draws 200-1000 milliwatts/megapixel. Here we show that radical reductions in electronic SWaP may be achieved using novel read-out and intial image processing architectures. Traditional architectures focus on "edge" analysis and compression of video data streams. Current standards use discrete cosine or wavelet transformations, followed by coefficient analysis and thresholding. Although cameras most typically use application specific circuits with hardware specifically designed for video compression, initial image processing, color demosaicing and compression still typically draws 5-10x more power per pixel than image capture. The strategies discussed in this paper can initially be implemented off-sensor to make the ISP power draw substantially less than the sensor operating power. Over the slightly longer term, implemetnation of compressive coding directly in the sensor read-out could reduce camera head power per pixel by as much as 100x by effectively eliminating the ISP while making the sensor itself 10x more power efficient. 

 Compressive sampling is an alternative to traditional compression strategies. 
 The work of Donoho, Candes and Tao showing that quasi-random global sampling kernels could inverted with high probability \cite{donoho, candes} 
 led many any studies of compressive sampling to focus on this approach \cite{eldar_kutyniok_2012, chartrand2010introduction}. However, in practice global random sampling is both hard to implement physically and mathematically unattractive, since the nonnegative nature of typical optical signals renders the forward model extremely ill-conditioned. Beyond its tremendous success in magnetic resonance imaging \cite{lustig2008compressed}, the most practical demonstrations of compressed sensing to date have been in tomographic systems where complete image sampling is not possible. Compressed sensing has shown clear advantages in projection, diffraction and spectral tomography\cite{compressTomo}, but in these cases the sampling kernel is as compact as possible. 

 More recently, neural methods have replaced constrained optimization as the inversion method of choice for compressively sampled data \cite{mousavi2017learning, lucas2018using, jin2017deep, kulkarni2016reconnet, xie2017adaptive}. When using an neural network, the invertibility proofs associated with quasi-random sampling are no longer relevant. Image data is typically "locally correlated," meaning that the the mutual information of pixel data drops rapidly as the separation between pixels grows. This means that it is unlikely that the numeric value of two measurements drawn from distant parts of an image can be represented in fewer than two digital values. Since pixels in compact neighborhoods tend to be highly correlated, however, one expects that these regions can be accurately described by less than one digital value per pixel. 
 
 This paper considers "compressed sensing" in the basic sense of the term, e.g. refering to blind downsampling of image data to less than one measurement per pixel. We do not consider quasi-random global sampling in the Donoho-Candes sense, since we expect that such sampling to be noncompetitive with local compression for the reason just mentioned. Our particular focus is video compression with the goal of radically reducing the size, weight and power per pixel of cameras. We specifically propose compressed sampling consisting of blind coded downsampling of pixel data with low-bit-depth-integer masks. "Blind" is the important aspect of compression here because any image data specific algorithm necessarily involves computational analysis of pixel values. Since each computation on the pixels costs power, blind compression tends to require significantly less power per pixel.
 
 While physical layer compressed sampling for video using, for example, multiple apertures~\cite{shankar,pitsianis} and coded pixel shape~\cite{brady2009coding} has been proposed, it has has not generally been applied in digital video sampling. Robucci {\it et al.} proposed compressive image sampling using quasi-random matrices in CMOS sensors~\cite{robucci2008compressive}, but with the goal of improved compression more paramount than power management. With no constraint on power, it is difficult for compressive measurement to compete with post-measurement compression. 
 Image and video compression is a highly developed technology centered on industrial standards such as JPEG, JPEG2000~\cite{jpeg}, H.264/AVC~\cite{itu264} and H.265/HEVC~\cite{itu265}. While computational complexity and computational power is an important consideration in defining such standards, the standards most commonly focus on processing image data at the point of encoding to preserve image features in easily decoded format. Typically, capture side processing is substantially more sophisticated than display side processing. This approach makes sense because image data is captured once but potentially displayed many times. One generally assumes that captured data will be displayed at least one time, making the expenditure of energy to prepare for standard decoding worthwhile. 
In many modern and emerging imaging applications, however, energy expended at the point of capture is more expensive than cloud or display energy. Capture devices are often mobile battery-powered systems with limited energy. 

Of course, numerous recent studies have applied variational autoencoders and related neural systems to image and video compression \cite{balle2018variational, mentzer2019practical, agustsson2018generative, minnen2018joint, lu2019dvc}. In many cases these systems achieve compression ratios that exceed HVEC and JPEG2000 quality. Unfortunately, due the rapidly developing state of this field we do not include direct image quality comparisons with our methods here. We anticipate, in fact, that some of the generative startegies developed in this field will be applicable to improving our results. To our knowledge, however, existing and previous work does not seek to minimize power in the compression step. We suggest, however, that the blind compressed sampling strategies proposed here are necessary for power minimization. More advanced neural estimation strategies may then apply unlimited computational power on off-camera platforms. 

Power management is particularly critical in view of the increasing popularity of  array cameras. Where one traditionally assumes that zoom, focus, exposure, etc. must be set at capture and that the captured image is processed and prepared for display at the camera head, parallel sampling allows the full ``light field'' of wide field of view, high resolution, high depth of field and high dynamic range image data to be captured. One can then filter at the cloud or display level to decide which subset of the light field to view. Since most of the light field is never viewed, however, analysis and processing of image data at the camera head is inefficient. For this reason, the compression methods discussed in this paper specifically relevant to array cameras, although they remain applicable to single camera systems. Array cameras \cite{parallel} aim to radically increase the quantity of camera image data through parallel optical and electronic processing. Since the quantity of captured information vastly exceeds the capacity of typical displays, most of the captured data is never viewed.  A  conventional camera assumes the image data will inevitably be displayed and thus it makes sense to process for display at the point of capture. An array camera, in contrast, may assume that the image data will never be displayed. In this case it makes sense to push most of the processing power requirement to the display side, since the odds that this power will be required are low. 

Compressed sensing, accordingly, removes computation from the camera head but increases the computation required at the end display. Of course, in video surveillance and other camera applications most of the data from even conventional single apeture cameras is never viewed and the techniques described here may be applicable. 

\subsection{Our contribution}  This report presents the DLACS method we develop, able to run both in the stand-alone manner and in the collaborative manner with other methods saving significant computation at the capture head and keeping good reconstruction quality, though requiring greater computation from a deep neural network on ISP of the display side. To our knowledge, this paper proposes for the first time blind coded downsampling of pixel data with low-bit-depth-integer masks. We show theoretically and experimental that this process uses 10-20x less power than JPEG compression. By delaying other traditional image processing steps, such as demosaicking, tone balancing, denoising, etc. for off camera or partial implementation, this strategy can reduce the power cost of imaging by upto 100x, which is critical for development of gigapixel scale array cameras.


\section{Coding strategy}

Conventionally image data is compressed according to ``intra-frame'' analysis and ``inter-frame'' prediction. Here we focus on the use of compressive sampling in intraframe compression to reduce camera head electrical power in array cameras. In conventional compression systems intraframe compression analyzes coefficients in discrete cosine or wavelet transformed data to code significant image features. As a practical matter, intraframe analysis describes the initial layer of data processing when image data captured at the sensor and transfered to the image signal processing (ISP) layer. In a conventional system the initial ISP layer converts captured data from raster scanned raw RGB data into image-based color planes in, for example, YUV, format. The initially formatted data is then further processed by a sequence of transformations to create compressed data in, for example, JPEG, H.264 or HEVC format. Each step of in the image processing pipeline can be described by linear transformations of pixel data in the form $g=Hf$, followed by nonlinear data analysis and filtering steps. Here $f$ refers to pixel data in any given layer and $g$ corresponds to coded data or ``measurements.'' In conventional systems the number of measurements is equal to or greater than the number of pixels. Meaning, for example, that when JPEG systems analyze the discrete Fourier transform of the frame every DCT coefficient is analyzed. In fact, conventional systems increase the number of measurements relative to the raw data in demosiacing RGB data prior to subsequent multi-step image data compression.

Compressive measurement systems also transform raw pixel data to measurements according to the linear transformation $g=Hf$, but in this case $H$ is of less than full rank. Compressive measurement implemented in digital form in the image read-out process yields the advantages of (1) immediately reducing that sensor data load and (2) reducing the number of times that each pixel value is independently processed. Since the power expended in the camera head ISP process is directly proportional to the number of pixels processed and the number of times each pixel value is accessed and processed, compressive measurement can substantially reduce ISP power. 

DLACS can be implemented in camera read-out as a few low-bit-depth-integer 2D masks. Image sensor data is read-out in row-frame format. Intra-frame compressive sampling is implemented by maintaining a set of buffers, each buffer persists for a few rows, here we consider buffers spanning 8, 16 or 32 rows. Each incoming pixel is added to the current state of one or more buffers, effecting in sum of pixel values weighted by the masks. Power and bandwidth is reduced in this process because (1) demosaicing and other early image processing tasks are deferred for later processing
, (2) the number of times each pixel is accessed and processed is much smaller than those in traditional methods, and (3) values in compressed file are not immediately analyzed for content. 

Traditional compressive sampling method usually use low-bit-depth-integer masks with certain properties, such as orthogonality, as weights for combining pixels in blocks of certain shapes. 
In this study, instead of manually making masks with orthogonal property, we use neural training methods to both obtaining the low-bit-depth-integer masks and training the weights of decompression layers. Based on characteristics of available system strategies, we used an integer code, the masks, to combine pixels of the raw-bayer data of each frame into compressed array of numbers. The compressed array of numbers are quantized to eight-bit integers as outputs of the compression module. The compressed data can be decompressed by a deep neural network and become the decompressed raw-bayer data. The usual RGB frame is obtained by applying the conventional steps of demosaicing, while-balance tuning, black-level tuning and color tuning to the decompressed raw-bayer data of the frame.

\begin{figure*}[htb]
\begin{center}
\includegraphics[width=0.5\textwidth]{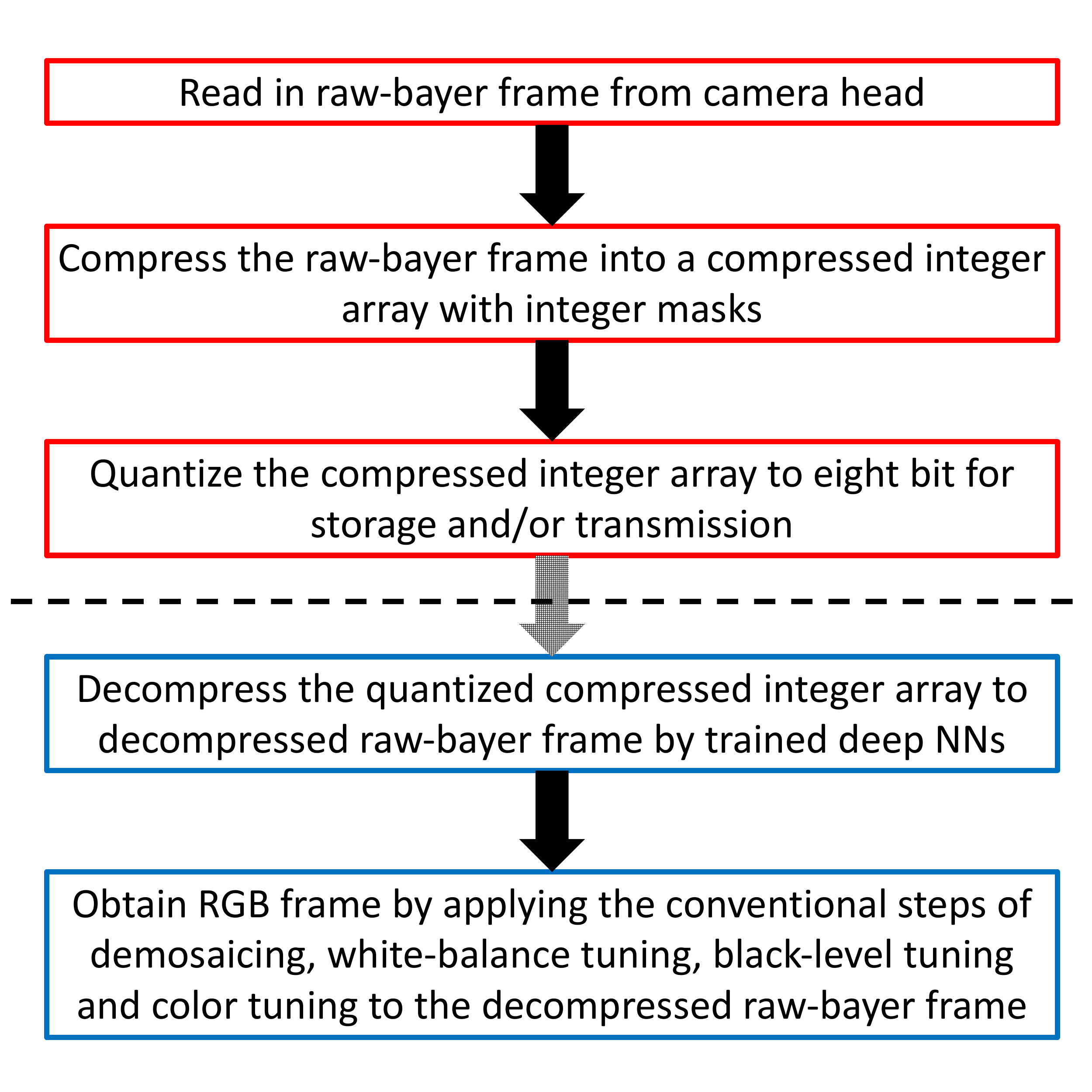}
\caption{\label{cs_flow} Flow of our DLACS method. The steps in the red boxes are carried out by the camera head and the electronics directly connected to the camera head. The quantized compressed data can be stored in and/or transmitted to local and/or cloud storage and/or buffer, as sketched by the shaded arrow crossed by a dashed line. The quantized compressed data can be decompressed and go through additional ISP steps by local and/or cloud devices independent from the camera-head system, as listed in the blue boxes.}
\end{center}
\end{figure*}

The flow of our DLACS method is presented in Fig.~\ref{cs_flow}. 
The steps carried out by the camera head and the electronics directly connected to it are listed in the red boxes of the flow.
The quantized compressed data from the camera-head system are stored in and/or transmitted to local and/or cloud storage and/or buffer.
Intermediate analyses can be carried out using the compressed data without decompression. Besides other analyses purposes, these analyses may help to determine whether the decompression and additional steps need to be processed.
When the images need to be displayed, or analyses of decompressed raw-bayer and/or RGB images need to be carried out, the compressed data can be decompressed by trained deep neural networks (NNs) and go through additional Image Signal Processing (ISP) steps in local and/or cloud devices independent from the camera-head system.

\begin{figure*}[htb]
\subfigure[Original]{
\begin{minipage}[t]{0.45\linewidth}
\centering
\includegraphics[width=0.9\textwidth]{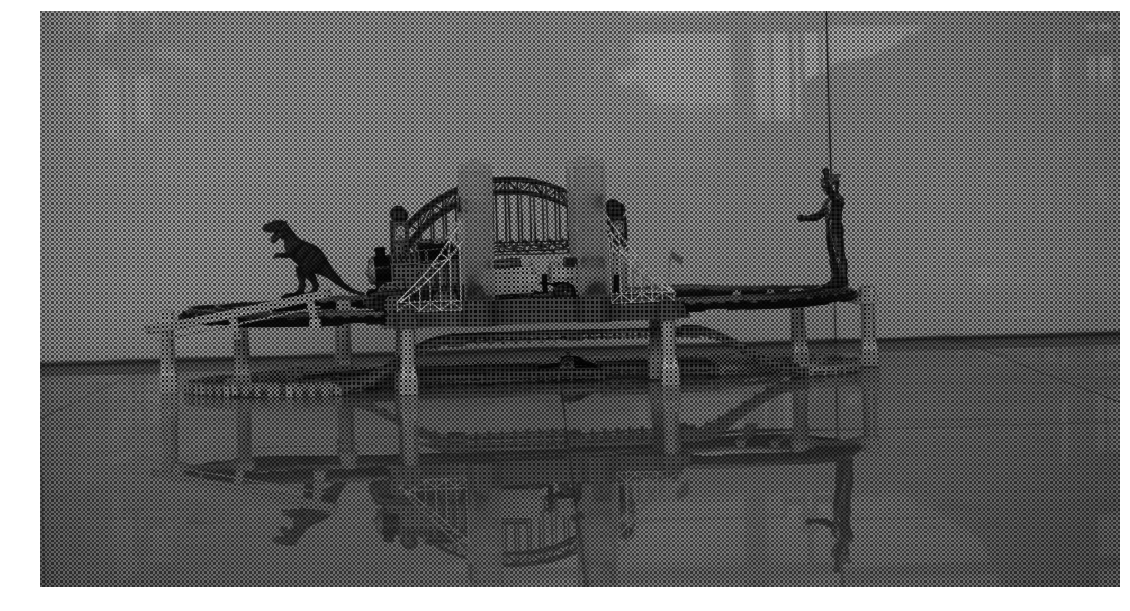}
\end{minipage}%
}%
\subfigure[Decompressed]{
\begin{minipage}[t]{0.45\linewidth}
\centering
\includegraphics[width=0.9\textwidth]{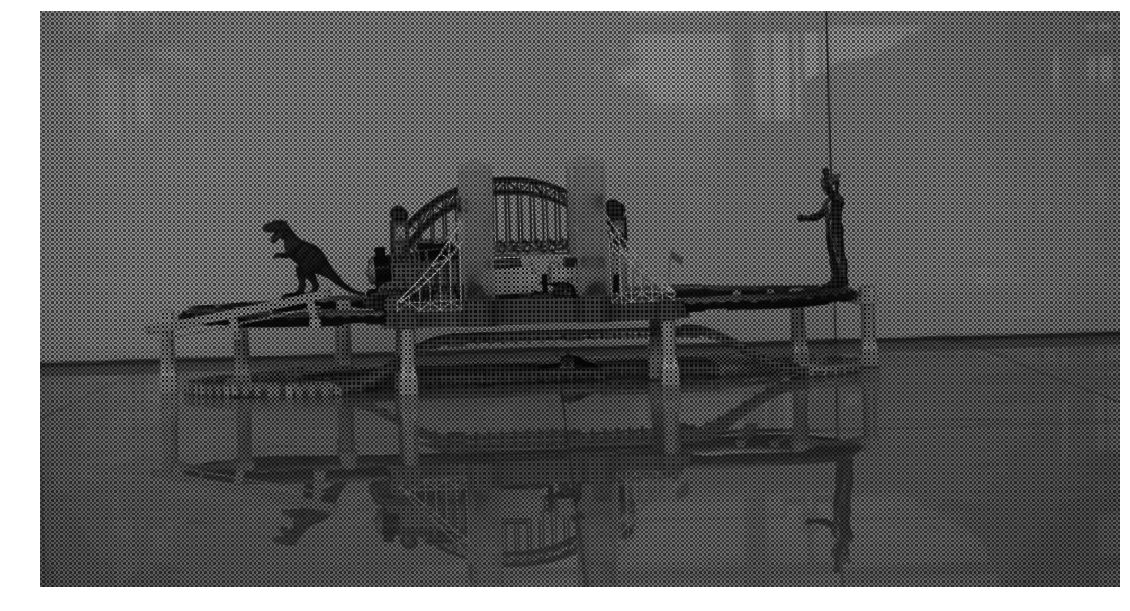}
\end{minipage}%
}%
\caption{\label{RBs} The original and the decompressed raw-bayer images are presented in the left and right panel, respectively.  The dimensions of both images are $[2048,\, 3864]$. The metrics of the decompressed relative to the original raw-bayer image are: $\textrm{PSNR} = 41.64$, $\textrm{SSIM} = 0.993$.}
\end{figure*}

\begin{figure*}[htb]
\includegraphics[width=0.9\textwidth]{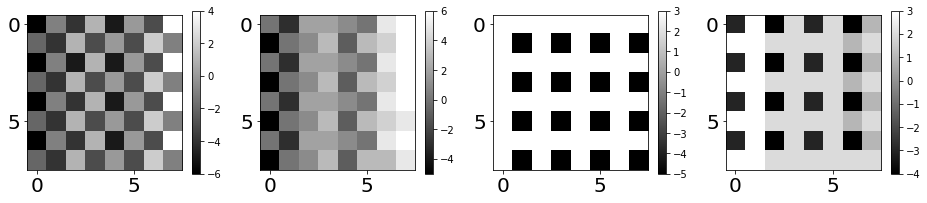}
\caption{\label{code_plot} Example four-bit-integer masks of dimension $[8,\, 8]$ in the four panels.}
\end{figure*}

\begin{figure*}[htb]
\includegraphics[width=0.9\textwidth]{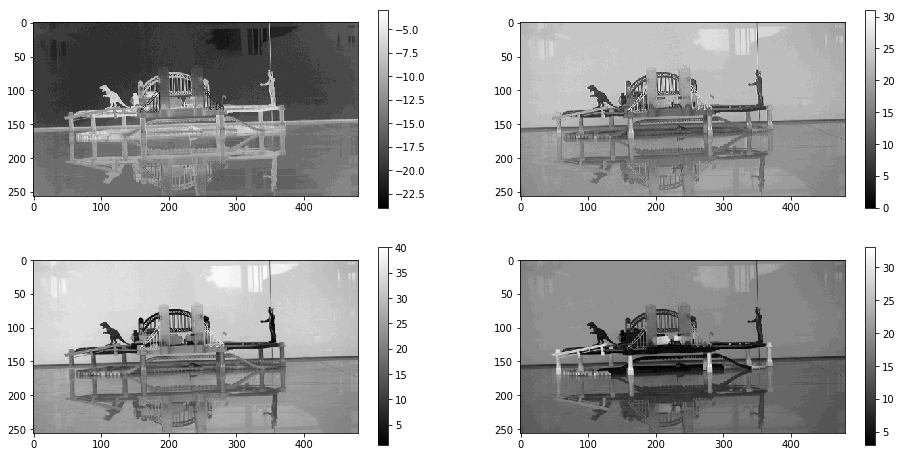}
\caption{\label{compressed_plot} The example compressed four eight-bit integer arrays of dimension $[256,\, 480]$ are visualized as four monochrome images in the four panels.}
\end{figure*}

An example raw-bayer frame from the camera head is presented in the left panel of Fig.~\ref{RBs}.
The raw-bayer frame from the camera head is immediately compressed by a set of four masks with integer values.
Example mask set containing four masks with four-bit integer values in dimension $[8,\, 8]$ are visualized in Fig.~\ref{code_plot}.
After compression, the compressed integer array is quantized to an eight-bit-integer array as presented in  Fig.~\ref{compressed_plot}. In proposed applications, these three operations occur in the camera head during the read-out process. Although we consider here only intra-frame compression, we recognize that these operations may be followed by loss-less entropy compression and/or intraframe compression as in current compression standards. 

\begin{figure*}[htb]
\subfigure[Original]{
\begin{minipage}[t]{0.45\linewidth}
\centering
\includegraphics[width=0.9\textwidth]{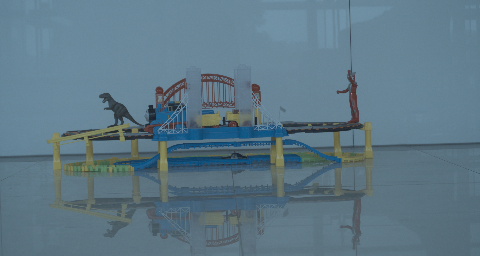}
\end{minipage}%
}%
\subfigure[Decompressed]{
\begin{minipage}[t]{0.45\linewidth}
\centering
\includegraphics[width=0.9\textwidth]{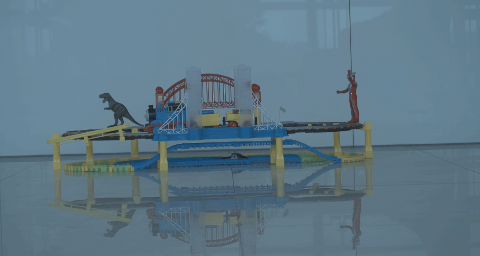}
\end{minipage}%
}%
\caption{\label{RGB_compare}(color online)  The RGB frames from the original and the decompressed raw-bayer frames are presented in the left and right panels, respectively. The metrics between these two are: $\textrm{PSNR} = 38.94$, $\textrm{SSIM} = 0.950$. The dimensions of both images are $[2048,\, 3864,\, 3]$.}
\end{figure*}

At the time of decompression, the eight-bit-integer array can be decompressed to a decompressed raw-bayer frame by trained deep NNs. The decompressed raw-bayer image from the eight-bit-integer array in Fig.~\ref{compressed_plot} is presented in the right panel of Fig.~\ref{RBs}.
The RGB frame can be obtained by applying the conventional steps of demosaicing, white-balance tuning, black-level tuning and color tuning to the decompressed raw-bayer frame. The RGB frames from the example original and decompressed raw-bayer data are presented in Fig.~\ref{RGB_compare}.

The processes of obtaining the integer code, the compression, the quantization of compressed data and the decompression, as well as the training of the neural-network (NN) system are described in the following paragraphs.

The compressive sampling operations, summing pixels weighted by masks, can be described by a specially designed convolutional-2D operation with a multi-channel kernel. The kernel with dimension $[k_x,\, k_y,\, 1, \, n_c]$ contains $n_c$ 2D masks with dimension $[k_x,\, k_y]$. Carrying out convolutional-2D operation using this kernel with stride size equal to kernel size is equivalent to using $n_c$ masks as weights in the summation of each blocks of pixels into a number in a no-gap-no-overlap manner.
A raw-bayer frame, $\textrm{Data}_{\textrm{in}}$ of dimension $[N_x,\, N_y, \, 1]$ is compressed into an array, $Comp_{\textrm{learn}}$, of dimension $[N_x/k_x, \, N_y/k_y, \, n_c]$.
A convolutional-2D-transpose layer of kernel size $[k_x,\, k_y,\, 1, n_c]$ and stride size $[k_x,\, k_y]$ decompresses $Comp_{\textrm{learn}}$ into $Decomp_{\textrm{learn}}$ which is in dimension $[N_x,\, N_y, \, 1]$.
The kernels of both layers are tuned via the training process which minimizes the mean-square-error (MSE) between $\textrm{Data}_{\textrm{in}}$ and $Decomp_{\textrm{learn}}$.
After training the masks, put together as $Comp_W$, are in float32 format, after being scaled by a constant, $sc_W$, they are rounded into integers of a certain bit depth, $Comp_{W\textrm{code}}$. The constant $sc_W$ is chosen so that the MSE between $Comp_W \cdot sc_W$ and $Comp_{W\textrm{code}}$ is minimized.
The masks are obtained so except for differences due to integerization of the values, they are optimized for combining pixels together in compressed arrays which can be decompressed by a single-layer convolutional-2D-transpose with highest quality.


The compressed array from using masks to combine raw-bayer pixels is scaled by a constant integer, $Q_{scale}$, and cast into an eight-bit-integer array, $Comp_Q$. The constant, $Q_{scale}$, is chosen so that the MSE between $Comp \cdot Q_{scale}$ and $Comp_Q$ is minimized when tested with the training data. The eight-bit-integer array, $Comp_Q$, is the output of the compression module.
Because the bit lengths of $Comp_Q$ and usual image format, such as eight-bit PNG and JPEG, are the same and raw-bayer data contains three channels of colors with the same amount of pixels, the compression ratio with respect to the non-compressed three-color eight-bit frame is $n_c/(3 \, k_x \, k_y)$.

\begin{figure*}[htb]
\includegraphics[width=0.9\textwidth]{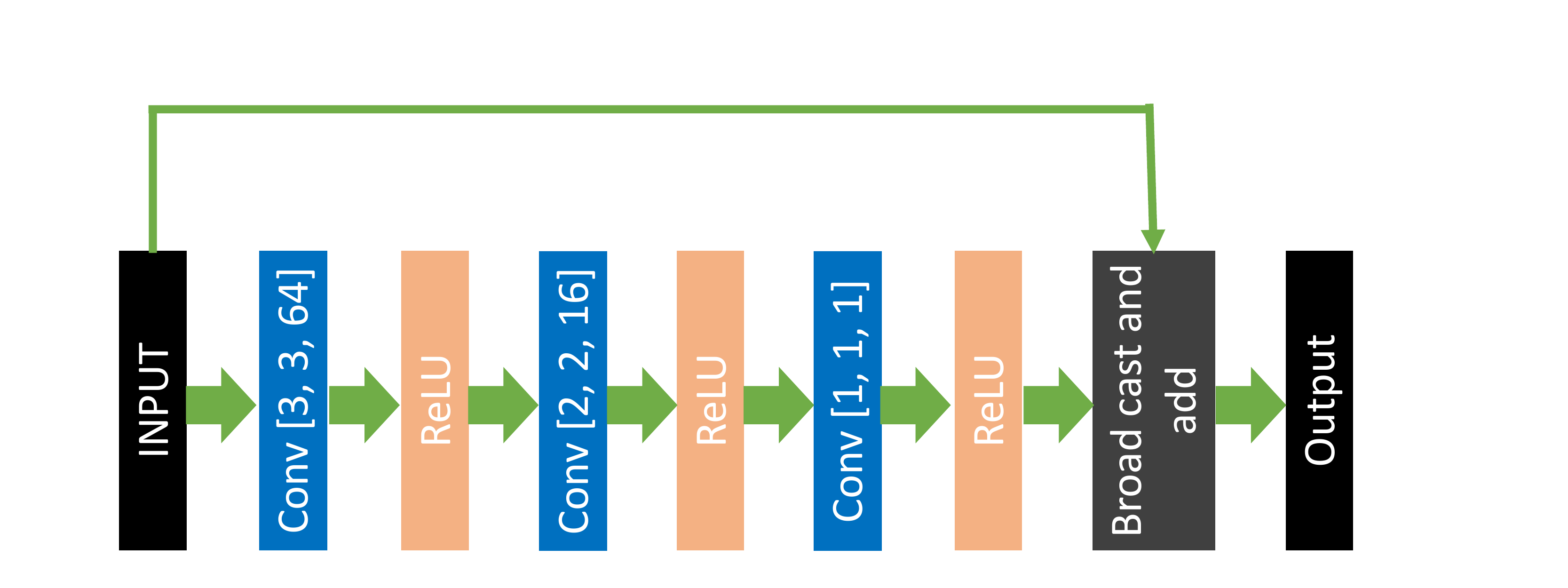}
\caption{\label{pd_NN}(color online).  Pre-transpose NN structure: Conv $[k_x,\, k_y,\, n_c]$ represents the convolution-2D layer with a kernel of dimension $[k_x,\, k_y,\, n_c]$, stride equal to 1 and same padding, ReLU represents the rectified-linear-unit activation. The output of the third ReLU with $n_c = 1$ is broad-casted to the same $n_c$ of the input and added to the input, forming a residual NN block.}
\end{figure*}
\begin{figure*}[htb]
\includegraphics[width=0.9\textwidth]{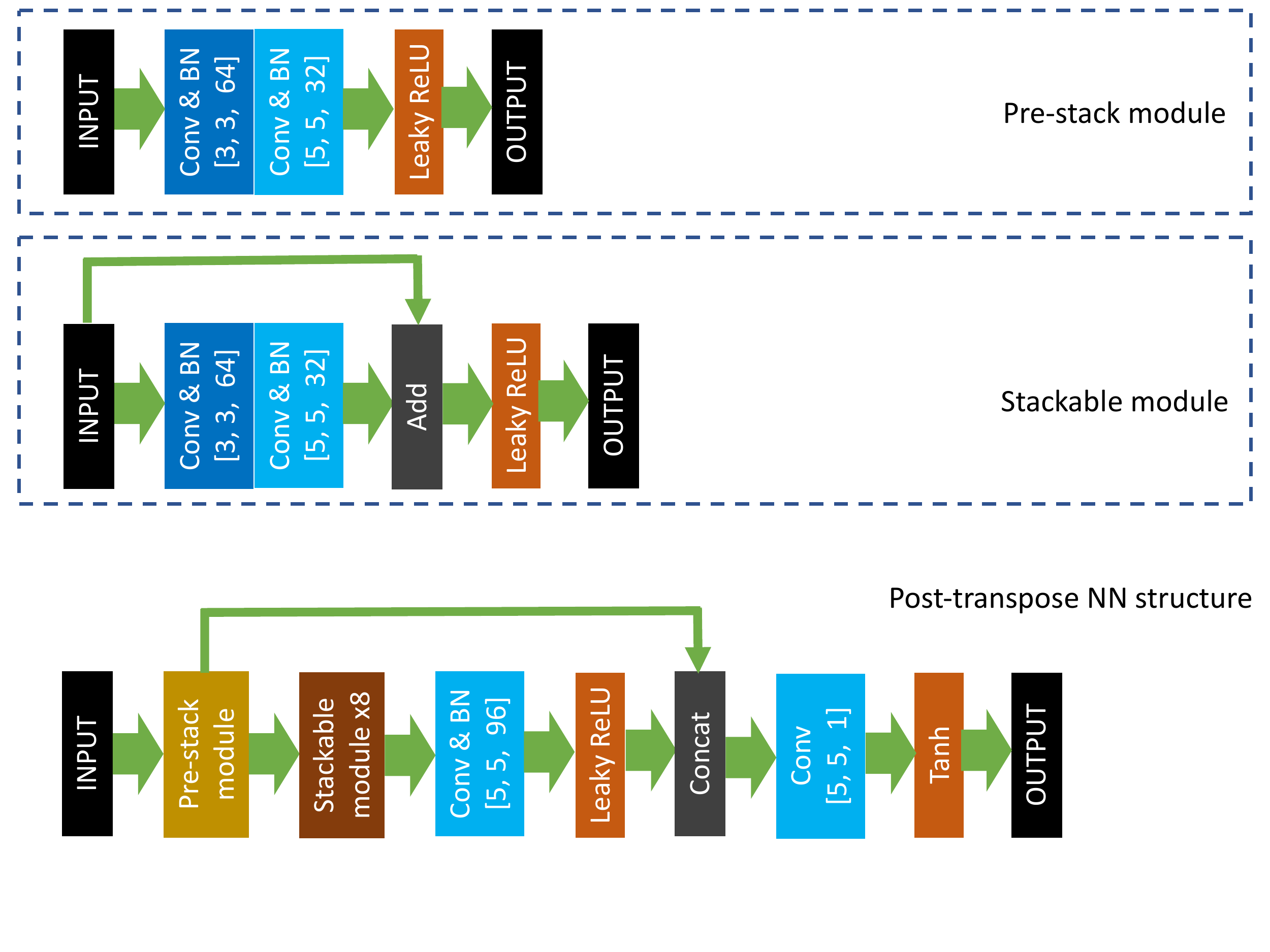}
\caption{\label{pp_NN}(color online).  Post-transpose NN: Conv represents a convolution-2D layer, the convolutional kernel dimension is presented in square brackets as $[k_x,\, k_y,\, n_c]$, BN represents Batch normalization, Leaky ReLU represents the leaky-ReLU activation, Concat represents concatenating two tensors in the channel dimension, and Tanh represents the hyperbolic-tangent activation. The pre-stack module is presented in the top panel, and the stackable module is presented in the middle panel. The post-transpose NN made of the pre-stack module, eight stackable modules stacked together, and a few additional layers is presented in the bottom panel.}
\end{figure*}

The eight-bit-integer array, $Comp_Q$, is the input of the decompression module, and is decompressed to be the decompressed raw-bayer data. In the decompression module, a pre-transpose  NN and a post-transpose NN are connected before and after the convolutional-2D-transpose layer.
The pre-transpose NN is a three-layer residual convolutional-2D NN, and its output, $dcomp_{\textrm{pre}}$, has the same dimension as $Comp_Q$. This NN can be conceptually viewed as compensating the loss due to the quantization process, and paves the way for the following steps of decompression. This NN is illustrated in Fig.~\ref{pd_NN}.
The convolutional-2D-transpose layer transforms $dcomp_{\textrm{pre}}$ into $dcomp$ which has the same dimension as the raw-bayer frame, $\textrm{Data}_{\textrm{in}}$.
The post-transpose NN is made of 20 convolutional-2D layers, 19 Batch-Normalization units and 10 Leaky-Relu activation units. Its output, $dcomp_{\textrm{pos}}$, has the same dimension as its input $dcomp$.  This NN can be viewed as carrying out post-processing to improve the quality of the output from the convolutional-2D-transpose layer, and its structure is illustrated in Fig.~\ref{pp_NN}.

The NN variables that need to be trained are the weight parameters in the decompression module because the compression module has only fixed parameters, the masks and the quantization-scaling constant.
The training data are $[128, 128, 1]$ crops of 4K raw-bayer images taken using the Mantis camera platform previously described in~\cite{parallel}.
The Mean-Square Error (MSE) between input and decompressed raw-bayer data is being optimized in the training process.

The DLACS method can also be applied on RGB data. In this case, the specially-designed convolutional-2D operation with masks contained in kernel $[k_x, \, k_y, \, 1,\, n_c]$ is applied to each of the channel of 2D dimension $[N_x, \, N_y]$. The compression ratio in this case is $n_c/(k_x k_y)$. Variations from this channel-by-channel approach of compressing RGB data can also be simply achieved by changing the kernel dimension and apply the convolutional operation on multiple channels at the same time.
Applying the CS method on RGB data involves more computational consumption compared with directly applying on the raw-bayer data when the sizes of output are the same: the raw-bayer data from the camera head need to be demosaiced into RGB first, which consumes computation, and because of the increase of the number of pixels, the amount of computation in the convolutional operation also increases.

The DLACS method can be conveniently and efficiently combined with existing image and/or video compression algorithms, such as entropy coding (EC), JPEG, JPEG2000, H.264 and HEVC.
A raw-bayer image can be compressed to an eight-bit integer array $Comp_Q$ of dimension $[N_x/k_x, \, N_y/k_y, \, n_c]$, which may be viewed as $n_c$ monochrome images of shape $[N_x/k_x, \, N_y/k_y]$. These monochrome images are of substantially smaller size than the original raw-bayer image and/or the RGB image demosaiced from the original raw-bayer image, and they can be further compressed by existing image-compression methods. After data storage and/or transmission, the raw-bayer image can be reconstructed by decompressing the reconstructed $Comp_Q$, where the reconstructed $Comp_Q$ can be obtained by using the decompression method corresponding to the existing method used to compress  $Comp_Q$.
The reconstructed raw-bayer image can then be demosaiced to RGB.
For video compression, a sequence of raw-bayer images can be compressed to a sequence of integer arrays $Comp_Q$s, and use existing video-compression algorithms to compress these integer arrays.
The reconstructed $Comp_Q$ of each frame can be decompressed and demosaiced.

The decompression NN can be re-trained or fine tuned (by carrying out transfer learning) for specific combinations with existing compression algorithms.
When combining with JPEG, the JPEG-compressed-and-reconstructed $Comp_Q$s are the input of the decompression NNs. The MSE between the output of the decompression, $dcomp_{\textrm{pos}}$, and the input raw-bayer data is minimized.
When combined with other existing algorithms, the NNs can be trained similarly.
Transformations between signed and unsigned eight-bit integers by adding or subtracting 128 need to be carried out to suit the data type required by the existing methods and the input and output data format of the NNs.




\subsection{Performance metrics}
As a demonstration of the performance of our DLACS method, three original raw-bayer images containing indoor and outdoor scenes are compressed by three sets of four masks. In the three sets,  masks are of dimensions $[8,\, 8]$, $[16,\, 16]$ and $[32,\, 32]$, and compression ratios due to DLACS encoding are $1/48$, $1/192$ and $1/768$, respectively. The RGB images are demosaiced from the reconstructed raw-bayer data and compared with compressed-and-decompressed results using other methods.
We use a well-established software, OpenCV, for downsampling and upsampling the RGB images demosaiced from the original raw-bayer images. The RGB images downsampled by OpenCV with proper parameters can reach desired compression ratios. The downsampled images can be upsampled to the original dimensions for measuring metrics with respect to the original RGB images.
We use a well-established software, Glymur, for JPEG2000 compression, and the desired compression ratio can be set manually and can be cross-checked by examining the input and output sizes.
We use a well-established software, Pillow, for JPEG compression. The desired compression ratio cannot be directly set but can be changed empirically by tuning the quality factor.
Compression ratio $1/48$ is reached by manually tuning the quality factors for different input images. For the three example images, the quality factors are between $70$ and $80$.
JPEG cannot reach a compression ratio as low as $1/192$ even when setting the quality factor to the minimal value, 1. The compression ratios reached by setting quality factor at 2 is close to those when setting quality factor at 1, but the quality is higher. In these comparisons, the quality factor is fixed at 2, and the compression ratios for the three example images are slightly larger than but close to $1/192$.
The comparisons of metrics using DLACS, down-and-up sampling (DAUS), JP2K and JPEG with compression  different ratios are summarized in Table.~\ref{table_48}.

\begin{figure*}[htb]
\subfigure[No compression.]{
\begin{minipage}[t]{0.3\linewidth}
\centering
\includegraphics[width=0.9\textwidth]{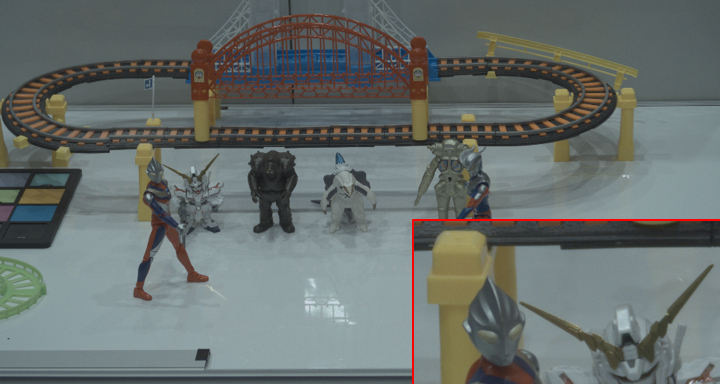}
\end{minipage}%
}%
\subfigure[$\textrm{Ratio} = 1/48$, $\textrm{PSNR} = 37.30$, $\textrm{SSIM} = 0.924$.]{
\begin{minipage}[t]{0.3\linewidth}
\centering
\includegraphics[width=0.9\textwidth]{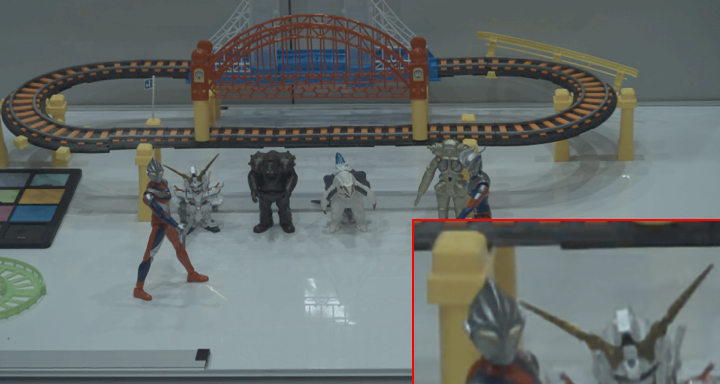}
\end{minipage}%
}%
\subfigure[$\textrm{Ratio} = 1/720$, $\textrm{PSNR} = 33.02$, $\textrm{SSIM} = 0.886$.]{
\begin{minipage}[t]{0.3\linewidth}
\centering
\includegraphics[width=0.9\textwidth]{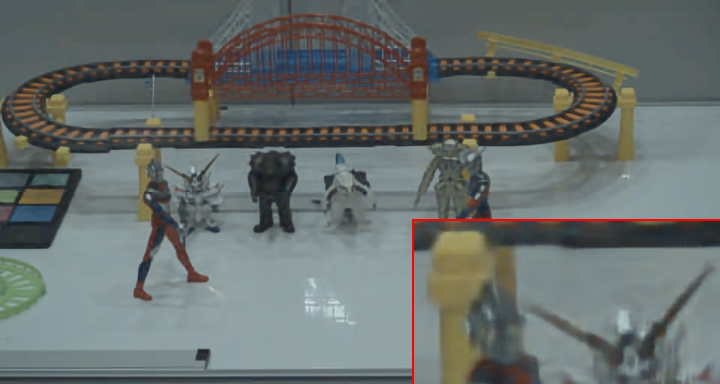}
\end{minipage}%
}%

\subfigure[No compression.]{
\begin{minipage}[t]{0.3\linewidth}
\centering
\includegraphics[width=0.9\textwidth]{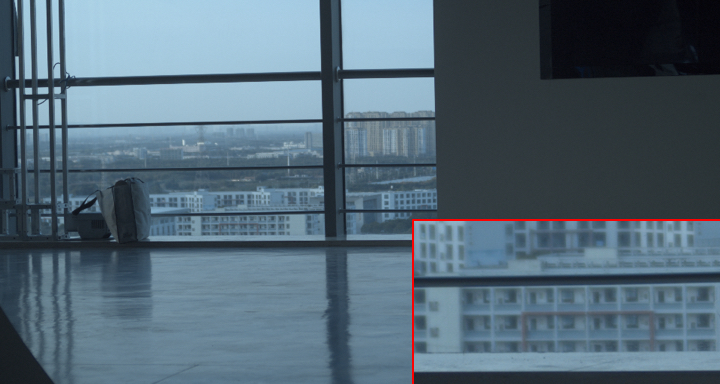}
\end{minipage}%
}%
\subfigure[$\textrm{Ratio} = 1/48$, $\textrm{PSNR} = 38.66$, $\textrm{SSIM} = 0.943$.]{
\begin{minipage}[t]{0.3\linewidth}
\centering
\includegraphics[width=0.9\textwidth]{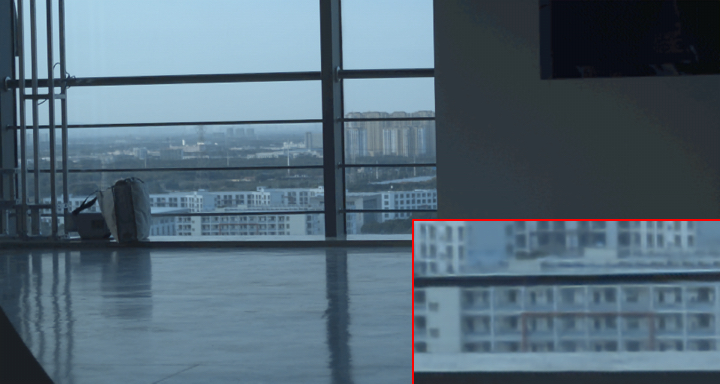}
\end{minipage}%
}%
\subfigure[$\textrm{Ratio} = 1/816$, $\textrm{PSNR} = 35.04$, $\textrm{SSIM} = 0.919$.]{
\begin{minipage}[t]{0.3\linewidth}
\centering
\includegraphics[width=0.9\textwidth]{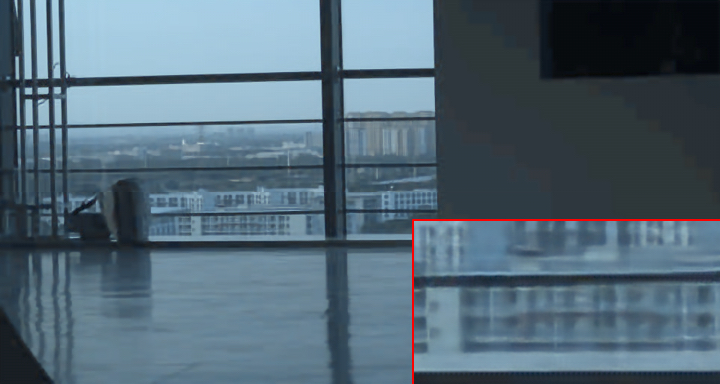}
\end{minipage}%
}%

\subfigure[No compression.]{
\begin{minipage}[t]{0.3\linewidth}
\centering
\includegraphics[width=0.9\textwidth]{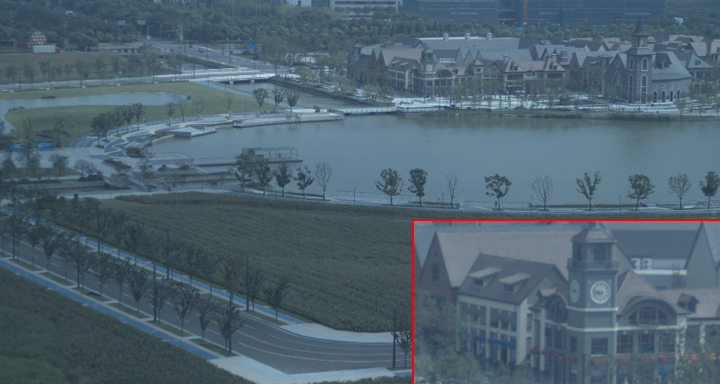}
\end{minipage}%
}%
\subfigure[$\textrm{Ratio} = 1/48$, $\textrm{PSNR} = 35.98$, $\textrm{SSIM} = 0.893$.]{
\begin{minipage}[t]{0.3\linewidth}
\centering
\includegraphics[width=0.9\textwidth]{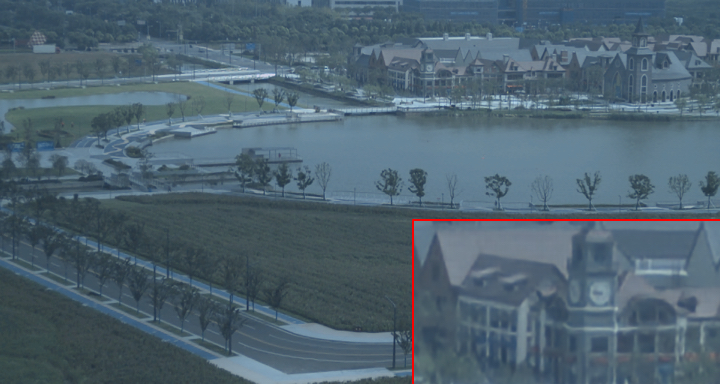}
\end{minipage}%
}%
\subfigure[$\textrm{Ratio} = 1/672$, $\textrm{PSNR} = 32.30$, $\textrm{SSIM} = 0.835$.]{
\begin{minipage}[t]{0.3\linewidth}
\centering
\includegraphics[width=0.9\textwidth]{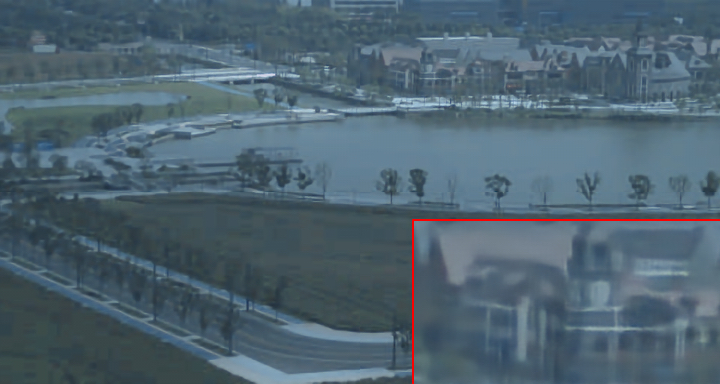}
\end{minipage}%
}%
\caption{\label{3_comp_scene_1_2_3}(color online).  Comparison of full 4K RGB images with different compression level for an indoor, a mixed-indoor-outdoor and an outdoor scenes are presented in the three rows of panels. The left panels present the original RGB images demosaiced from the original raw-bayer data. The middle panels present the RGB images demosaiced from reconstructed raw-bayer data compressed by a four masks of dimension $[8,\, 8]$. The right panels present the RGB images of this hybrid method combining DLACS with four masks of dimension $[8,\, 8]$ and JPEG with $\textrm{Q} = 85$. The compression ratios and metric relative to the original RGB images are presented in the panels. Parts of the full images are zoomed in and presented in the red boxes on the bottom right of each panel to demonstrate the reconstruction-quality drop as more compression are carried out.}
\end{figure*}

Our DLACS method can be easily combined with other methods. 
As an example, the compressed eight-bit integer arrays $Comp_Q$s are further compressed by arithmetic coding, which is one of the well-established method of EC. When EC is applied, $Comp_Q$ is compressed to a smaller size in a lossless manner. The comparison of quality between DLACS + EC, JPEG and JP2K at the same level of compression are presented in Table~\ref{CS_EC_RB_123}.
As another example, the compressed eight-bit integer arrays $Comp_Q$s are further compressed by JPEG with a fixed quality-factor $\textrm{Q}$, and the reconstructed $Comp_Q$s from the JPEG files are decompressed and demosaiced. The decompression NNs are trained for decompressing the reconstructed $Comp_Q$s.
The comparisons of the RGB images from the original raw-bayer, the reconstructed raw-bayer data under compression ratio $1/48$ and from the reconstructed raw-bayer data from the DLACS-JPEG combined method are presented in Fig.~\ref{3_comp_scene_1_2_3}.
The drop of reconstruction quality, as described by the metrics, related to DLACS method compressing at $1/48$ ratio and the hybrid-DLACS-JPEG method compressing at ratios between $1/624$ to $1/1104$ are clearly visualized in the zoomed-in windows on the bottom right of the panels.
While the hybrid-DLACS-JPEG method removed more fine details compared with the original and the $1/48$ compressed-and-reconstructed images, the results still keep high quality as can be visualized in the figures and observed in the metrics.
The code corresponding to Tables~\ref{table_48} and \ref{CS_EC_RB_123}, and Fig.~\ref{3_comp_scene_1_2_3} can be found in Reference~\cite{CS_RB_code}.

It is observed in these examples, the reconstruction quality of the stand-alone DLACS method is lower than that of JPEG2000's in general, lower than JPEG with compression ratio $1/48$, close to JPEG with compression ratio between $1/90$ and $1/125$ with the help from EC, and higher than JPEG with compression ratio $1/192$. 
While JPEG cannot reach a compression ratio below $1/192$, the stand-alone DLACS method and the hybrid-DLACS-JPEG method can reach much smaller compression ratio with reasonable quality. 
And because of the substantially reduced amount of computation by the stand-alone DLACS method and the hybrid-DLACS-JPEG method, they can reach small compression ratios with reasonable quality consuming much less computation and power in the compression module.

\begin{table}[hbt]
\begin{center}
\begin{tabular}{|c|c|c|c|c|c|c|}\hline
    & \multicolumn{2}{|c|}{Image 1} & \multicolumn{2}{|c|}{Image 2} & \multicolumn{2}{|c|}{Image 3} \\ \cline{2-7}
    & PSNR & SSIM & PSNR & SSIM & PSNR & SSIM \\ \hline
DLACS ($1/48$) & $37.30$ & $0.924$ &  $38.66$ & $0.943$ & $35.98$ & $0.893$ \\ \hline
DAUS ($1/48$) & $35.07$ & $0.912$  & $35.05$ & $0.935$ & $33.86$ & $0.873$ \\ \hline 
JP2K ($1/48$) & $41.69$ & $0.955$ & $43.28$ & $0.965$ & $40.84$ & $0.951$ \\ \hline
JPEG ($1/48$) & $41.33$ & $0.953$  & $43.39$ & $0.967$ & $39.83$ & $0.944$ \\ \hline
DLACS ($1/192$) & $32.60$ & $0.881$ & $33.84$ & $0.913$ & $32.17$ & $0.836$ \\ \hline
DAUS ($1/192$) & $30.19$ & $0.868$  & $30.10$ & $0.907$ & $30.16$ & $0.823$ \\ \hline 
JP2K ($1/192$) & $39.25$ & $0.945$ & $41.20$ & $0.960$ & $37.53$ & $0.917$ \\ \hline
JPEG ($1/183$) & $25.69$ & $0.817$  & $23.59$ & $0.815$ & $23.02$ & $0.730$ \\ \hline
DLACS ($1/768$) & $28.29$ & $0.842$ & $29.00$ & $0.876$ & $29.17$ & $0.800$ \\ \hline
DAUS ($1/768$) & $26.32$ & $0.837$ & $25.42$ & $0.881$ & $27.29$ & $0.797$ \\ \hline 
JP2K ($1/768$) & $35.22$ & $0.916$ & $38.06$ & $0.948$ & $34.34$ & $0.873$ \\ \hline
\end{tabular}
\end{center}
\caption{Comparison of results with different compression ratios. The compression ratios are presented in the brackets in the leftmost column.}
\label{table_48}
\end{table}

\begin{table}[hbt]
\begin{center}
\begin{tabular}{|c|c|c|c|c|c|c|}\hline
    & \multicolumn{2}{|c|}{Image 1} & \multicolumn{2}{|c|}{Image 2} & \multicolumn{2}{|c|}{Image 3} \\ \cline{2-7}
    & Ratio & SSIM & Ratio & SSIM & Ratio & SSIM \\ \hline
DLACS ($1/48$) + EC & $1/94$ & $0.924$ & $1/95$ & $0.943$ & $1/102$ & $0.893$ \\ \hline
JPEG & $1/94$ & $0.938$ & $1/95$ & $0.960$ & $1/102$ & $0.899$ \\ \hline
JP2K & $1/94$ & $0.950$ & $1/95$ & $0.962$ & $1/102$ & $0.935$ \\ \hline
DLACS ($1/192$) + EC & $1/320$ & $0.881$ & $1/330$ & $0.913$ & $1/363$ & $0.836$ \\ \hline
JP2K & $1/320$ & $0.938$ & $1/330$ & $0.957$ & $1/364$ & $0.898$ \\ \hline
DLACS ($1/768$) + EC & $1/1335$ & $0.842$ & $1/1373$ & $0.876$ & $1/1584$ & $0.800$ \\ \hline
JP2K & $1/1337$ & $0.897$ & $1/1371$ & $0.938$ & $1/1574$ & $0.848$ \\ \hline
\end{tabular}
\end{center}
\caption{Comparison of SSIM between DLACS with EC, JPEG and JP2K for the three scenes.}
\label{CS_EC_RB_123}
\end{table}

\begin{figure*}[htb]
\subfigure[No compression.]{
\begin{minipage}[t]{0.3\linewidth}
\centering
\includegraphics[width=0.9\textwidth]{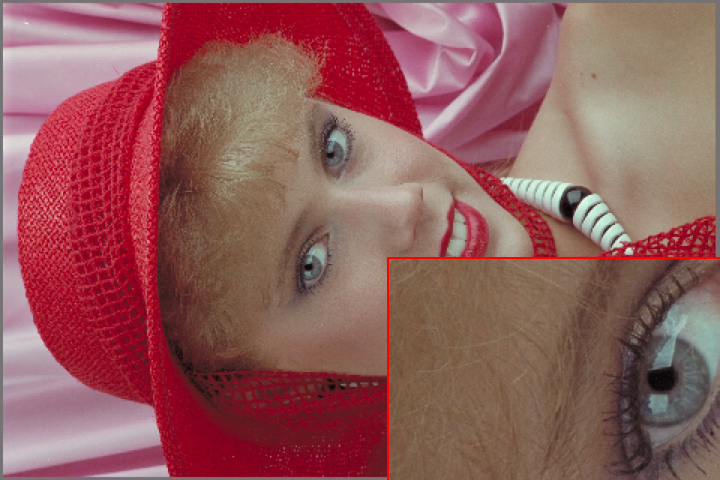}
\end{minipage}%
}%
\subfigure[$\textrm{Ratio} = 1/16$, $\textrm{PSNR} = 36.29$, $\textrm{SSIM} = 0.906$.]{
\begin{minipage}[t]{0.3\linewidth}
\centering
\includegraphics[width=0.9\textwidth]{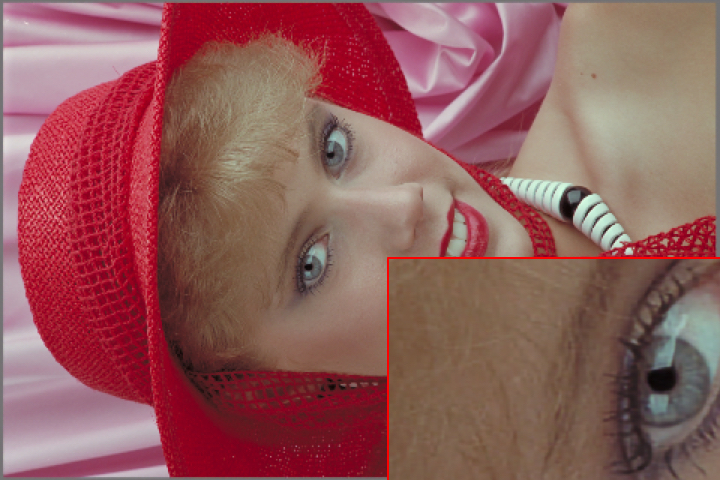}
\end{minipage}%
}%
\subfigure[$\textrm{Ratio} = 1/256$, $\textrm{PSNR} = 28.51$, $\textrm{SSIM} = 0.752$.]{
\begin{minipage}[t]{0.3\linewidth}
\centering
\includegraphics[width=0.9\textwidth]{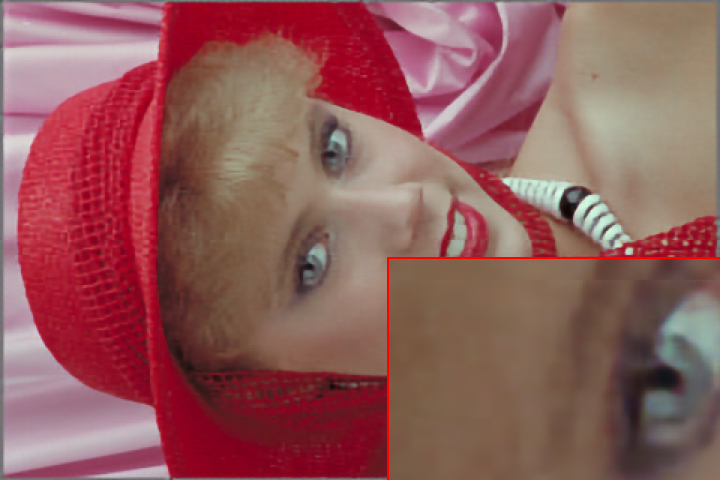}
\end{minipage}%
}%

\subfigure[No compression.]{
\begin{minipage}[t]{0.3\linewidth}
\centering
\includegraphics[width=0.9\textwidth]{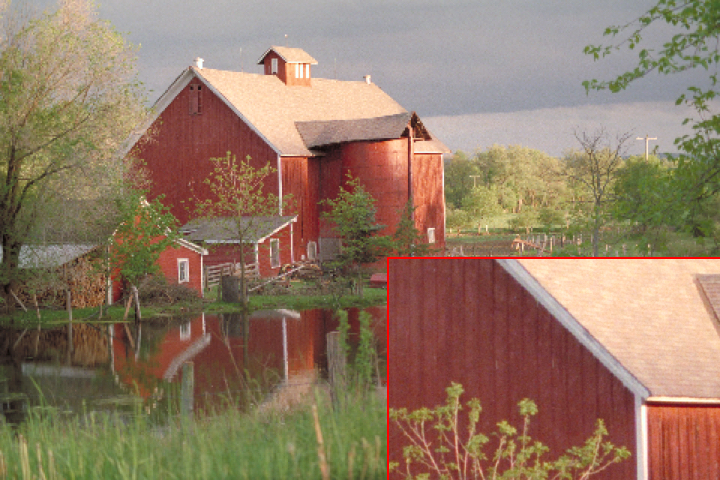}
\end{minipage}%
}%
\subfigure[$\textrm{Ratio} = 1/16$, $\textrm{PSNR} = 34.84$, $\textrm{SSIM} = 0.883$.]{
\begin{minipage}[t]{0.3\linewidth}
\centering
\includegraphics[width=0.9\textwidth]{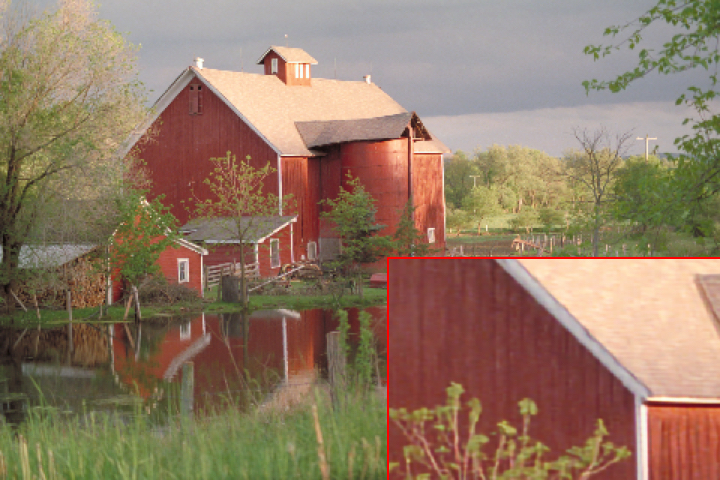}
\end{minipage}%
}%
\subfigure[$\textrm{Ratio} = 1/256$, $\textrm{PSNR} = 27.38$, $\textrm{SSIM} = 0.689$.]{
\begin{minipage}[t]{0.3\linewidth}
\centering
\includegraphics[width=0.9\textwidth]{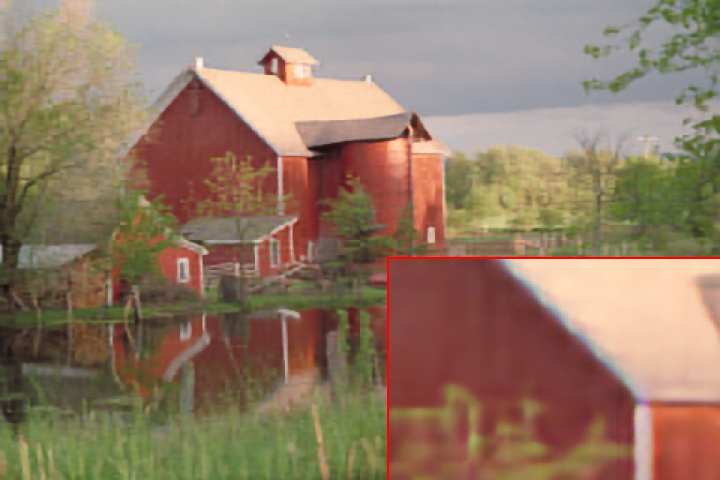}
\end{minipage}%
}%

\subfigure[No compression.]{
\begin{minipage}[t]{0.3\linewidth}
\centering
\includegraphics[width=0.9\textwidth]{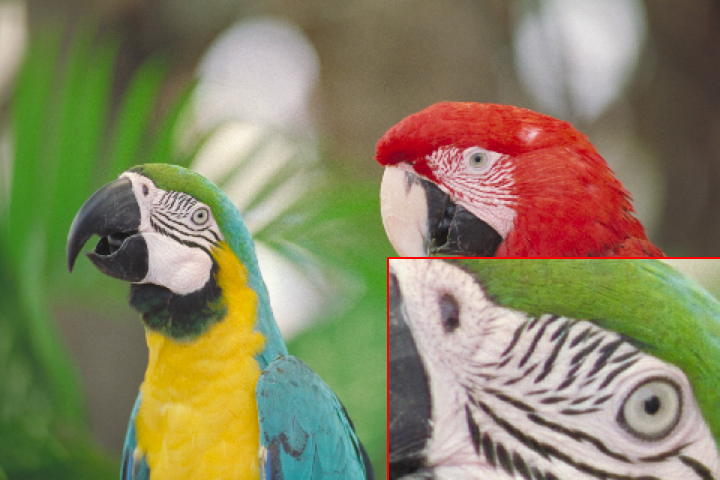}
\end{minipage}%
}%
\subfigure[$\textrm{Ratio} = 1/16$, $\textrm{PSNR} = 39.39$, $\textrm{SSIM} = 0.928$.]{
\begin{minipage}[t]{0.3\linewidth}
\centering
\includegraphics[width=0.9\textwidth]{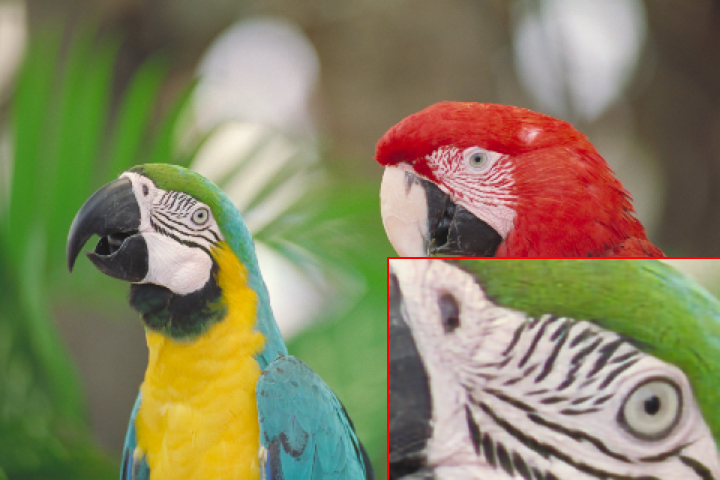}
\end{minipage}%
}%
\subfigure[$\textrm{Ratio} = 1/256$, $\textrm{PSNR} = 31.11$, $\textrm{SSIM} = 0.858$.]{
\begin{minipage}[t]{0.3\linewidth}
\centering
\includegraphics[width=0.9\textwidth]{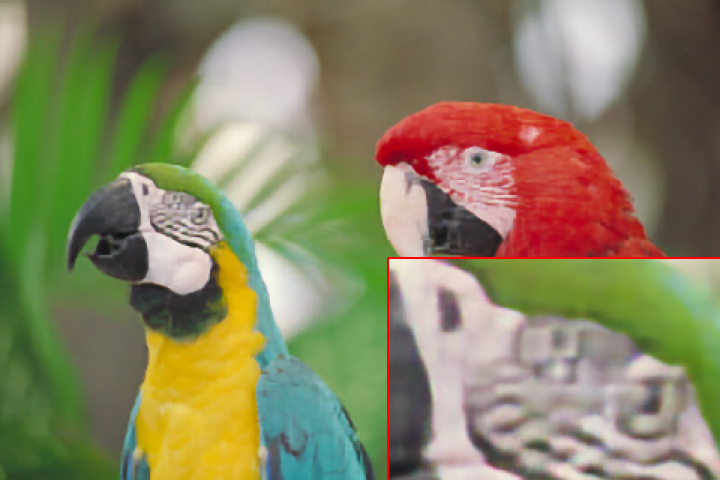}
\end{minipage}%
}%
\caption{\label{KD_RGBs}(color online).  Comparison of three Kodak RGB images with different compression levels are presented. The left panels present the original Kodak RGB images. The middle panels present the reconstructed RGB images after being compressed by kernel of dimension $[8,\, 8,\, 1, 4]$. The right panels present the reconstructed RGB images after being compressed by kernel $[32,\, 32,\, 1, 4]$. The compression ratios and metric relative to the original RGB images are presented in the panels. Parts of the full images are zoomed in and presented in the red boxes on the bottom right of each panel to demonstrate the reconstruction-quality drop as more compression are carried out.}
\end{figure*}

Our DLACS method is also tested using the uncompressed Kodak dataset in RGB, in a channel-by-channel manner: each of the R, G and B channel of dimension $[N_x, \, N_y]$ is independently compressed and decompressed in the same manner as when compressing/decompressing the raw-bayer data.
After decompression of each channel, instead of demosaicing, the three reconstructed channels are combined into RGB images.
The quality of reconstruction is measured by comparing the original and reconstructed RGB images.
The comparison of three original Kodak RGB images with corresponding reconstructed ones using only the CS method under compression ratio of $1/16$ and $1/256$, as well as the metrics, are presented in Fig.~\ref{KD_RGBs}.
The quality comparison between only DLACS, DLACS + EC, JPEG and JP2K are presented in Tables.~\ref{Kodak_cs16}, \ref{Kodak_cs64} and \ref{Kodak_cs256}.
Similar to the DLACS on raw-bayer study above, at same compression levels, DLACS without EC generally has lower quality compared with JPEG and JP2K, DLACS + EC has close quality to JPEG but lower than JP2K, and JPEG cannot reach a compression ratio below $\approx 1/185$.

The code corresponding to Tables~\ref{Kodak_cs16}, \ref{Kodak_cs64} and \ref{Kodak_cs256}, and Fig.~\ref{KD_RGBs} can be found in Reference~\cite{CS_Kodak_code}.


\begin{table}[hbt]
\begin{center}
\begin{tabular}{|c|c|c|c|c|c|c|}\hline
    & \multicolumn{2}{|c|}{Kodak 1} & \multicolumn{2}{|c|}{Kodak 2} & \multicolumn{2}{|c|}{Kodak 3} \\ \cline{2-7}
    & Ratio & SSIM & Ratio & SSIM & Ratio & SSIM \\ \hline
DLACS (no EC) & $1/16$ & $0.906$ & $1/16$ & $0.883$ & $1/16$ & $0.928$ \\ \hline
JPEG & $1/16$ & $0.948$ & $1/16$ & $0.943$ & $1/16$ & $0.955$ \\ \hline
JP2K & $1/16$ & $0.957$ & $1/16$ & $0.958$ & $1/16$ & $0.963$ \\ \hline
DLACS + EC & $1/25$ & $0.906$ & $1/26$ & $0.883$ & $1/23$ & $0.928$ \\ \hline
JPEG & $1/25$ & $0.933$ & $1/26$ & $0.925$ & $1/23$ & $0.946$ \\ \hline
JP2K & $1/25$ & $0.943$ & $1/26$ & $0.938$ & $1/23$ & $0.955$ \\ \hline
\end{tabular}
\end{center}
\caption{Comparison of quality between DLACS with/without EC, JPEG and JP2K. Four masks of dimension $[8,\, 8]$ for each channel of RGB, achieving DLACS-only compression ratio of $1/16$.}
\label{Kodak_cs16}
\end{table}


\begin{table}[hbt]
\begin{center}
\begin{tabular}{|c|c|c|c|c|c|c|}\hline
    & \multicolumn{2}{|c|}{Kodak 1} & \multicolumn{2}{|c|}{Kodak 2} & \multicolumn{2}{|c|}{Kodak 3} \\ \cline{2-7}
    & Ratio & SSIM & Ratio & SSIM & Ratio & SSIM \\ \hline
DLACS (no EC) & $1/64$ & $0.822$ & $1/64$ & $0.773$ & $1/64$ & $0.894$ \\ \hline
JPEG & $1/64$ & $0.878$ & $1/64$ & $0.863$ & $1/64$ & $0.912$ \\ \hline
JP2K & $1/64$ & $0.914$ & $1/64$ & $0.896$ & $1/64$ & $0.929$ \\ \hline
DLACS + EC & $1/93$ & $0.822$ & $1/93$ & $0.773$ & $1/89$ & $0.894$ \\ \hline
JPEG & $1/93$ & $0.840$ & $1/93$ & $0.818$ & $1/89$ & $0.896$ \\ \hline
JP2K & $1/93$ & $0.914$ & $1/93$ & $0.875$ & $1/89$ & $0.922$ \\ \hline
\end{tabular}
\end{center}
\caption{Comparison of quality between DLACS with/without EC, JPEG and JP2K. Four masks of dimension $[16,\, 16]$ for each channel of RGB, achieving DLACS-only compression ratio of $1/64$.}
\label{Kodak_cs64}
\end{table}


\begin{table}[hbt]
\begin{center}
\begin{tabular}{|c|c|c|c|c|c|c|}\hline
    & \multicolumn{2}{|c|}{Kodak 1} & \multicolumn{2}{|c|}{Kodak 2} & \multicolumn{2}{|c|}{Kodak 3} \\ \cline{2-7}
    & Ratio & SSIM & Ratio & SSIM & Ratio & SSIM \\ \hline
DLACS (no EC) & $1/256$ & $0.752$ & $1/256$ & $0.689$ & $1/256$ & $0.858$ \\ \hline
JPEG & $1/177$ & $0.640$ & $1/176$ & $0.619$ & $1/181$ & $0.745$ \\ \hline
JP2K & $1/256$ & $0.848$ & $1/256$ & $0.816$ & $1/256$ & $0.905$ \\ \hline
DLACS + EC & $1/355$ & $0.752$ & $1/359$ & $0.689$ & $1/346$ & $0.858$ \\ \hline
JP2K & $1/355$ & $0.830$ & $1/359$ & $0.792$ & $1/346$ & $0.899$ \\ \hline
\end{tabular}
\end{center}
\caption{Comparison of quality between DLACS with/without EC, JPEG and JP2K. Four masks of dimension $[32,\, 32]$ for each channel of RGB, achieving DLACS-only compression ratio of $1/256$.}
\label{Kodak_cs256}
\end{table}

\subsection{Computational complexity}
In the compression process, the DLACS method with set-of-four masks of 2D-mask-dimensions $[8,\, 8]$, $[16,\, 16]$ and $[32,\, 32]$ consists of four integer-integer multiplications and four integer-integer additions when combining pixels with masks, and one integer-integer division for quantization (which may be further simplified to bit shifting) for each raw-bayer pixel.
The standard JPEG method \cite{JPEG_standard} carries out Discrete Cosine Transformations (DCTs) on $8 \times 8$ blocks of YCbCr pixels which are demosaiced and color transferred from the raw-bayer data. The DCT step of JPEG involves 64 integer-float-number multiplications and 64 float-float-number additions for each Y, Cb and Cr pixel. The quantization step of JPEG involves one time float-integer-number multiplication.
Comparing the convolutional operation of the CS method and the DCT operation of JPEG, the JPEG method involves $(64/4) \times 3 = 48$  times more multiplications and additions, where $3$ comes from the number of channels of the YCbCr format. 

We carried out simple tests in Python to confirm the relative  computational complexity of the encoding operations used for DLACS and the DCT used for JPEG. In these tests DCT processing required 22 times the computation time per pixel.
Both the DLACS-encoding operation and the DCT are coded as functions with loops on a CPU without any acceleration methods such as vectorization or parallel running.
The run time is calculated only when the CPU is running the loops while the data transmission and/or storage and/or display time are not included.
The type of all the data involved in the computation are fixed as 64-bit float, to avoid subtle changes of run time due to data-type transformations.
These tests do not include a fundamental-level optimization for operations on data with different bit depth, which can be carried out on specific hardware with careful design.
The code of these tests can be found in Reference~\cite{Conv_vs_DCT_code}.

We also carried out running-speed comparison between DLACS-encoder and DCT using CUDA (C++) codes, optimized to the same level for parallel running on Nvidia-TX1 system.
Six 4K camera heads are connected to our Nvidia-TX1 system. Using a common data acquisition tool, V4L2, 4K raw-bayer frames can be taken by specified camera heads and be stored in the main memory of the TX1 GPU.
Each raw-bayer frame in the main memory is transferred to the CUDA buffer of the GPU for desired computation.
After CUDA computation, the results in CUDA buffer are transferred to the main memory, and then can be written to the hard-disk or transferred out via certain interface of the TX1 board.
For comparing running speed of the DLACS-encoder and DCT in a fair manner, the tests are carried out while making sure the capacity of the TX1 GPU is exhausted.
To keep the GPU's capacity exhausted throughout a test, and CUDA not waiting for data coming to the main memory, we  store a number of prepared 4K raw-bayer frames, the DLACS masks and the DCT coefficients in the main memory prior to running CUDA, enabling CUDA to start processing the next frame right after finishing the current one.
Some initial runs have been carried out, saving output to the hard-disk of TX1, for verifying values in output file being correct for DLACS-encoder and DCT computation.
After verification, we carried out long runs without saving the output file to the hard-disk, and potential effects on running time results due to CUDA waiting for hard-disk writing of computation results have been avoided. 
The running-time values are saved to the hard-disk, and this process takes negligible time and has negligible effect on CUDA running time.
In addition, we record time of computation and time of transferring between CUDA buffer and the main memory separately, and keep the comparison only on the time of computation.
With 10 parallel CUDA threads, cross checked by using a few different amounts of threads, we made sure each of the test exhausted the computation capacity of the Nvidia-TX1 GPU. 
Results of computation time with low statistical uncertainties are obtained in the long runs with multiple parallel CUDA threads, and are presented in Table.~\ref{CUDA_CS_vs_DCT}.
It is found that the computation time per pixel from DCT ($8 \times 8$ block) is more than 14 times of that from DLACS-encoder (set-of-four masks).
These comparisons yield ratio of computational complexity close to a simple estimation of $(64/4) = 16$.

\begin{table}[hbt]
\begin{center}
\begin{tabular}{|l|c|c|}\hline
Run type & Thread number & Running time per pixel (pico second) \\ \hline
DLACS, $[8,\, 8, 4]$ & 10 & $294.48 \pm 3.57$ \\ \hline
DLACS, $[16,\, 16, 4]$ & 10 & $295.51 \pm 3.15$ \\ \hline
DLACS $[32,\, 32,4]$ & 10 & $285.37 \pm 3.19$ \\ \hline
DCT & 10 & $4323.42 \pm 29.82$ \\ \hline
DLACS $[8,\, 8, 4]$ & 15 (for cross check) & $302.27 \pm 2.90$ \\ \hline
DCT & 12 (for cross check) & $4270.52 \pm 27.75$ \\ \hline
\end{tabular}
\end{center}
\caption{Running time per pixel of DLACS-encoding and DCT on Nvidia-TX1 system, using full computation capacity of GPU, with 10 parallel CUDA threads. Two cross-check runs with different amount of threads are also presented on the bottom of this table. The uncertainties in this table represent statistical uncertainty of each run. The three numbers in the square brackets for DLACS runs represent X and Y dimensions of masks and number of masks, respectively.}
\label{CUDA_CS_vs_DCT}
\end{table}

It should be noted that in the standards of 4:2:1 and 4:2:0, the Cb and Cr channels are recorded in dimensions different from the Y channel, due to the smaller 2D size of these channels, the theoretical ratio $48$, in comparison with JPEG, becomes $32$ and $24$, respectively.
In addition the bit depth of each operation in the JPEG method is larger than that in the DLACS method because float numbers require more bits than integers.
The decrease of bit depth could lead to different levels of decrease of computation depending on the hardware and the details of fundamental-level control of memory, buffer, etc. of the algorithm on the hardware.
When the same type of EC algorithm is being used, because of the decrease of the dimension of the integer arrays to be coded by EC, the EC step used for DLACS needs much less computation than that used in other methods such as JPEG and JP2K (the comparison between different EC algorithms is not in the scope of this study). 
Without considering the entropy-coding and the bit depth difference, the computational complexity of the DLACS method is at least $\approx 20$ times simpler than that of the JPEG method. In addition, the demosaicing process carried out on the raw-bayer data to produce the RGB and/or YCbCr data before JPEG and/or JP2K compression is also skipped in our DLACS method, which further reduces the amount of computation in the compression process.

In comparative studies of  JP2K and JPEG the computational complexity of compression (JP2K over JPEG) varied from $\approx 5$ to $8$ times under different tests in a ``fair comparison'' condition~\cite{CC_JPEG2000, SANTACRUZ2002113, CC_JPEG2000_1}, meaning that the DLACS methods proposed here reduce computaional complexity more than $100x$ relative to JP2K.

\section{Conclusion}
Digital cameras implement a pixel data processing pipeline with power requirements linearly proportional to the number of computational operations per pixel. By reducing this number by >20x, compressive sampling enables a >20x reduction in camera head processing power per pixel. DLACS pays for this reduction by requiring substantially greater display side image data processing, but in ultra-high resolution imaging systems most pixels are never examined and, for unexamined pixels, this cost is never paid. Even when it is paid, display side processing may occur on cloud platforms where processing power is more economical. 

By reducing bit depth in our DLACS method, the base-level computation design may be further simplified.
When running in the stand-alone manner, because the compression process only involves multiplication , summing and division of integers, the computation is significantly simpler than that of JPEG, JPEG2000.
When running in the combined manner, instead of the original raw-bayer data or the RGB data demosaiced from them, the integer arrays, $Comp_Q$s, serve as input of the existing algorithms. Because the size of $Comp_Q$s is substantially smaller, much less amount of computation need to be carried out by the existing algorithms.
For example, DLACS compressed data may also be conveniently combined with NNs for different purposes such as face recognition, etc.

The size, weight and power of electrical components is currently the primary barrier to compact gigapixel scale cameras. As described here, compressive sampling combined with deep learning based decompression can resolve this barrier.


\bibliographystyle{siamplain}
\bibliography{refs}











\end{document}